# Self-assembly of the chiral donor-acceptor molecule DCzDCN on Cu(100)

*Robert Ranecki, Benedikt Baumann, Stefan Lach, Christiane Ziegler*

RPTU Kaiserslautern-Landau, Department of Physics and Research Center OPTIMAS, Erwin-Schroedinger-Str. 56, 67663 Kaiserslautern (Germany)

**Abstract:** Donor-acceptor (D-A) structured molecules are essential components in organic electronics. The respective molecular structure of these molecules and their synthesis are primarily determined by the intended area of application. Typically, D-A molecules promote charge separation and transport in organic photovoltaics (OPV) or organic field-effect transistors (OFET). D-A molecules showing a larger twist angle between D and A units are, e.g., extremely important for the development of high internal quantum efficiency in organic light-emitting diodes (OLEDs). A prototypical molecule of this D-A type is DCzDCN (5-(4,6-diphenyl-1,3,5-triazin-2-yl)benzene-1,3-dinitrile). In most cases, these molecules are only investigated regarding their electronic and structural interaction in bulk aggregates but not in ultra-thin films supported by a metallic substrate. Here, we present growth and electronic structure studies of DCzDCN on a Cu(100) surface. In a complementary approach, through the use of Scanning Tunneling Microscopy and Spectroscopy (STM and STS), we were able to view both the adsorption geometry and the local electronic states of the adsorbed molecules in direct comparison with the integral electronic structure of the DCzDCN/CU(100) interface using Ultraviolet and Inverse Photoemission Spectroscopy (UPS and IPS). The orientation of the molecules with the donor part towards the substrate results in a chiral resolution at the interface due to the molecular as well as the substrate symmetry and additional strong molecular electrostatic forces. Thus, the formation of various bulk-unlike homochiral structures and the appearance of hybrid interface states (HIS) modifies the molecular electronic properties of the DCzDCN/Cu(100) system significantly compared to that of a single DCzDCN molecule. This may be not only useful for optoelectronic applications but also in organic spintronics.

Donor/acceptor (D-A) systems and molecules are among the most important building blocks in the field of molecular electronics. The choice of D-A-molecules and interfaces is very diverse, depending on the application. Especially for D-A systems used in optoelectronic devices, research ranges from nanostructured molecular heterojunctions [1-3] to systems with intramolecular charge transport [4-7] and D-A modifications by targeted interaction with a substrate [8]. Especially for the applications in highly effective light emitting diodes (OLED), special requirements are placed on the D-A systems used there [7,9]. Of particular interest are D-A systems that use reverse intersystem crossing (rISC) to massively increase fluorescence yield. Such so-called TADF molecules (thermally activated delayed fluorescence) are widely used in current high-efficiency OLED devices [10]. The DCzDCN molecule (Fig. 1) was designed as a universal host material for TADF OLEDs (organic light-emitting diodes) [11].

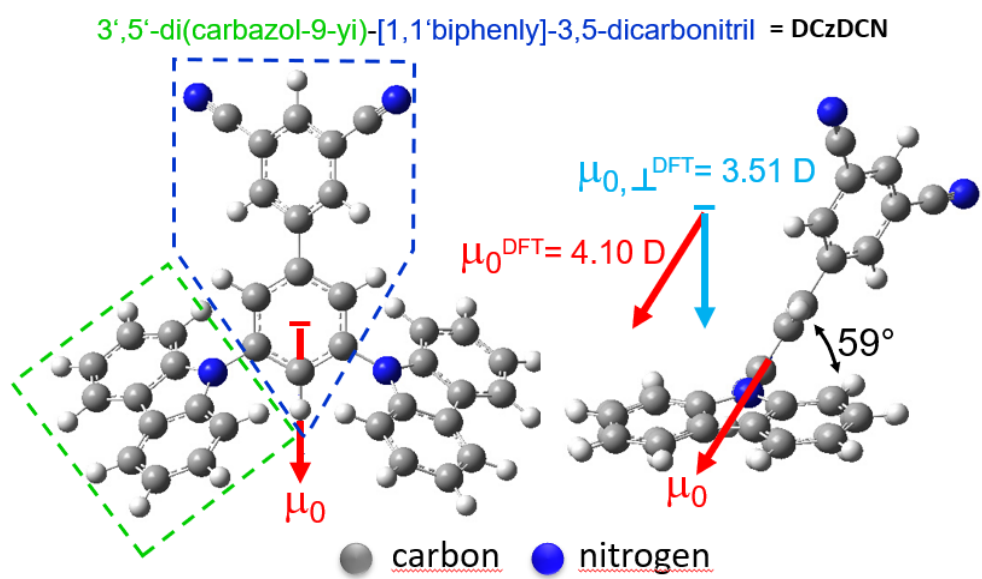

*Figure 1 Molecular structure of DCzDCN calculated by DFT. The phthalonitrile part (upper part in the blue dashed box) and the carbazole parts (green dashed box) are oriented in a tilted (59°) and twisted configuration. The calculated total molecular dipole of $\mu_0$ = 4.1 D and its perpendicular part of 3.51 D are shown.*

It combines a weak carbazole D-unit also acting as hole transport unit (green dashed box in Fig. 1) and a strong phthalonitrile A-unit for electron transport (upper part in the blue dashed box in Fig. 1). Both units are coupled via a twisted biphenyl link. Biphenyl compounds are rotamers by themselves, but in the DCzDCN, the substituents add axial chirality. Thus, two enantiomers are present in the bulk phase. Furthermore, the twisting of the D-A units against each other causes localization of the highest occupied molecular orbital (HOMO) on the D unit and the highest unoccupied molecular orbital (LUMO) on the A unit. This localization of the two frontier orbitals is the cause of the occurrence of an rISC in this molecule. The estimation of the energies of the two frontier orbitals has so far only been determined within the framework of cyclic voltammetry measurements (CV) [11]. Thereby, an electron affinity of -3.26 eV and an ionization potential of -6.14 was found. However, TADF-active substances are not only usable for OLEDs, but there are a lot more possible applications for these materials, like bioimaging probes or oxygen sensors [12,13]. It is also conceivable to exploit the rISC in the context of an organic spintronic application. Since many of these extended applications, outside the typical use in an OLED device, usually imply interaction with some kind of a supporting substrate, the interaction with a Cu(100) substrate was investigated in this study. Particularly in the case of molecules that are typically only used in thick layers, often little is known about their adsorption behavior. However, adsorption on a surface can lead to new molecular organization units, that are not necessarily predictable from the known structures in thick layers or the single molecule. This can be, for example, a surface template effect, which can lead to the modification of the intermolecular dipole interaction and thus of the opto-electronic properties [14]. It is therefore important to choose the broadest possible analytical approach for such systems. For this purpose, a complementary in situ approach by means of Scanning Tunneling Microscopy and Spectroscopy (STM and STS) and Ultraviolet and Inverse Photoemission Spectroscopy (UPS and IPS) was used. Here, in addition to the topographic information from the STM, particular attention was paid to the combined discussion of the single molecular spectroscopy received by the STS measurements and the integral information from the whole DCzDCN/Cu(100) interface by UPS/IPES. Using this approach, the detection of possible hybrid interface states (HIS) in the region around the Fermi level can also be spatially resolved on the molecule.

**Results**

**Scanning Tunneling Microscopy**

Fig. 2 a-d show constant current topography images obtained at 9.5 K of 0.25 ML - 0.9 ML of DCzDCN molecules deposited onto the Cu(100) surface kept at room temperature. From the UPS and IPES data for the 1 ML coverage, we were able to estimate that we need a relatively high bias voltage application to ensure that the deeply-in-energy-lying boundary orbitals will be involved in the tunneling processes. Therefore, we performed STM images at the relatively high bias voltage of ±3 V but with a low current setpoint of 50 pA to provide large enough tip-molecule separation.

At the early stage of growth (Fig. 2a), there is both a decoration of the terrace steps and a distributed adsorption on the large-scale (ranging up to 200 nm) atomically flat terraces. In the case of a negligible interaction of the molecules with the Cu substrate, only a preferential decoration of the edges and a resulting layer/island growth would occur. The additional formation of molecular clusters thus hints at a hybridization with the metallic surface, which will later be discussed in the UPS section.

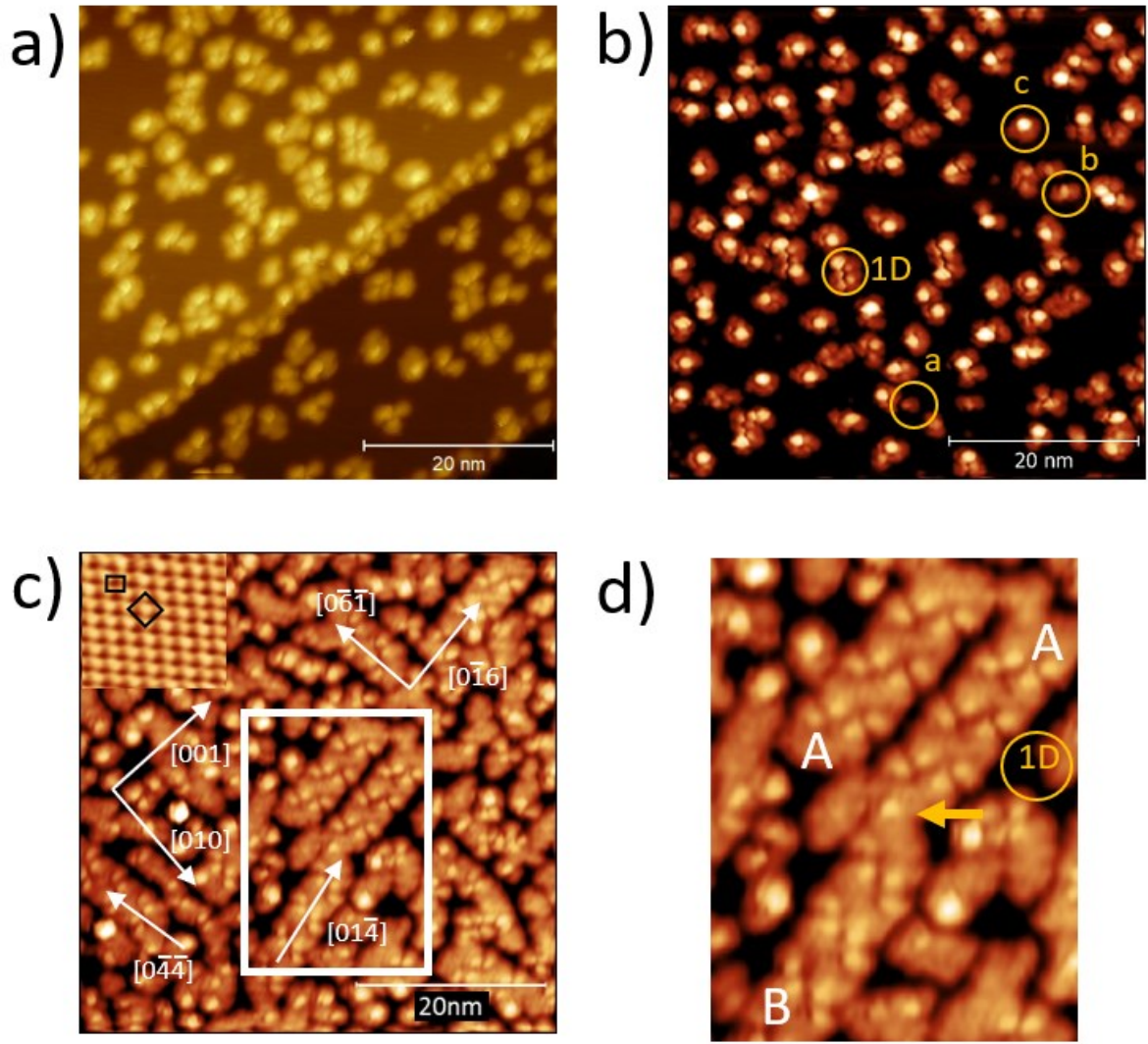

*Fig. 2 a-c) STM constant current topography image obtained at 9.5 K, 50 pA, for sub-monolayer DCzDCN coverage on Cu(100): a) 0.25 ML 50 nm × 50 nm (bias voltage 3 V), b) 0.25 ML 50 nm × 50 nm (bias voltage -3 V). Various recurring uniform molecular clusters are characterized with respect to their increasing size (a-1D). c) 0.9 ML showing molecular 1D self-assembly structures along two perpendicular crystalline directions (bias voltage 3 V). For comparison, the inset shows an atomically resolved STM image of the clean Cu(100) surface obtained at a bias voltage of 50 mV and a current setpoint of 300 pA. d) Zoom into the white frame of c), showing the "tire track"-like 1D structures with two different homochiral domain structures (A,B). The yellow arrow shows the transition between two such domains.*

Going into details of the STM contrast topography images, the comparison of Fig. 2a (3 V) and Fig. 2b with reverse bias (-3 V) shows that some parts of the molecular adsorbates are much more prominent (brighter) than the other parts. The detailed analysis of reverse bias (-3 V) topography images in Fig. 2b implies that the same parts of the molecular structure are more prominent for both polarities. Taking into account the molecular boundary orbitals from the DFT calculations for the occupied states (see supporting materials Fig. S1 orange area), a narrow energy range is found in which both the MOs with an exclusive localization on the two carbazole units and an MO with a localization on the phthalonitrile biphenyl moiety are found. In the selected energy range, units localized on the phthalonitrile biphenyl moiety are, in particular, dominant for the unoccupied states. Therefore, we attributed the brighter parts of the molecular structures to the phthalonitrile unit. This means the preferred adsorption geometry of the DCzDCN molecules is upright in a face-on position with respect to the carbazole units on the Cu(100) surface. Furthermore, in Fig. 2b four types of repeating structures can be distinguished (marked by a-c and 1D). As the coverage increases to 0.95 ML (Fig. 2c), the DCzDCN molecules form mainly "tire track"-like 1D structures, preferably growing in two directions perpendicular to each other. Furthermore, even at almost full ML coverage (Fig 2c), we can see that molecules still prefer to find free vacancies on the surface instead of starting a second layer, which is a clear indication of interaction with the substrate.

The high-resolution STM images of the cluster units identified in Fig. 2b shown in Fig. 3 indicate that structures a-c are, in fact, single molecules, dimers, and tetramers.

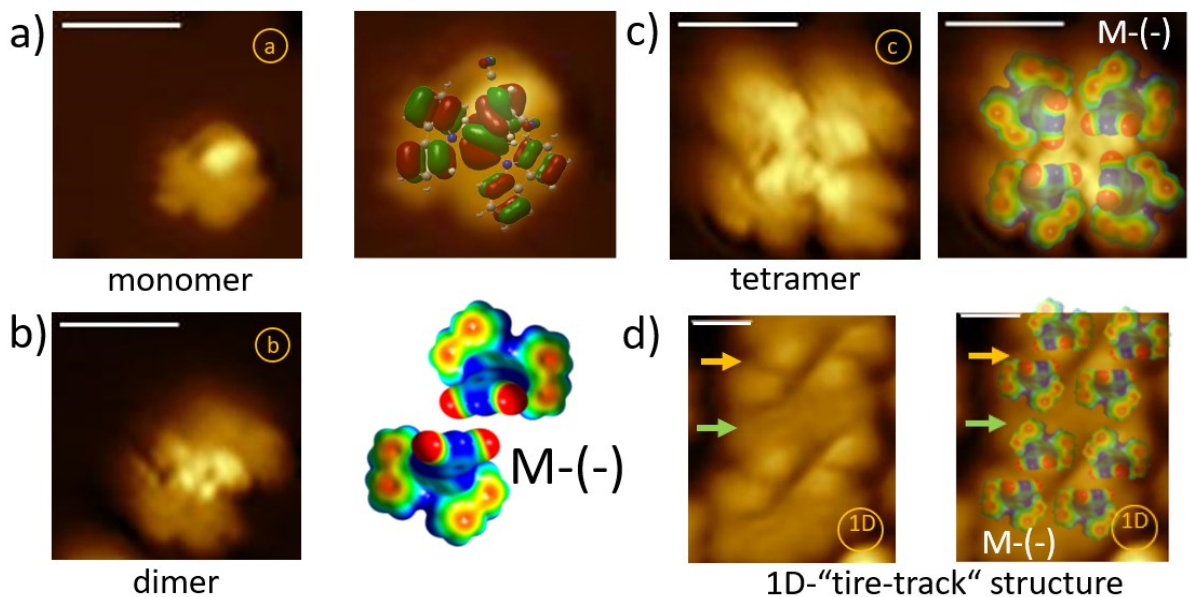

*Fig. 3 Submolecular resolution images of the different molecular clusters identified in Fig. 2. a) Single DCzDCN molecule. Shown on the right is the overlay of the image with the MOs in question for the bias voltage used. b) Dimer structure with M-(-) chirality. On the right, an electrostatic potential map of the dimer shows the centers of electrostatic interaction. c) Tetramer forming a homochriral M-(-) phase. On the right, overlay of the image with a tetramer model d) 1D homochiral (M-(-)) "tire track"-like structure formed by dimer pairs. On the right, overlay of the image with a staggered dimer model., The alternating flat carbazole (green arrow) area and the vertical phthalonitrile biphenyl moiety (yellow arrow) are clearly visible.*

As shown in Fig. 3a, submolecular resolution reveals the nature of the different molecule clusters in Fig. 2. Related to structure a in Fig. 2b, Fig. 3a shows a single DCzDCN molecule. The clear resolution of the respective double-ring structure of the carbazole units in direct contact with the substrate is evident. The angled (see Fig. 1) and, thus, eccentric structure of the phthalonitrile biphenyl moiety relative to the carbazole plane can also be seen very clearly in Fig. 3a. This submolecular structure is also consistent with the superposition of MOs expected in the bias region used, which illustrates the particularly exposed tunneling channel across the phthalonitrile group, as shown in Fig. 3a. Therefore, we assign the three structures to **a** = single molecule, **b** = dimer, and **c** = tetramer. All the structures found result from the strong electrostatic interactions between the vertical phthalonitrile biphenyl moieties. Such interactions can be nicely illustrated by the electrostatic potential map (ESP) of the single molecule (Fig. S2 supporting information and Fig. 3b - d ). The strong interaction of the negative nitrile groups (red color in the ESP representation) of one molecule with the positive aromatic phthalonitrile part (blue color in the ESP representation) of another molecule leads to a dimerization. This is also favored by the vertical angle of 59° of the phthalonitrile biphenyl moiety, then pointing to each other. The comparison of the molecule structure and the STM image of a dimer pair (Fig. 3b) also shows that only two identical rotamers (M-(-) here) form a dimer pair. The same mechanism also leads to the formation of the square tetramers in Fig. 3c. Also for the tetramers, a homochirality is maintained. At sub-ML coverage, in addition to a few clearly identifiable ensembles, dimeric and predominantly tetrameric structures can be found. The evaluation of the individual ensembles in Fig. 2b shows that 72% of all assignable structures are tetramers (see supporting information Fig. S3). This changes when a coverage near 1 ML is reached. Now, the tetramers are no longer the dominating structures. "Tire track"-like 1D structures (Fig. 2c+d and Fig. 3d) are formed. This "tire track" structure is built of staggered dimer pairs, as shown in the overlayer with the ESP model in Fig. 3d. For each of the two enantiomers, two homochiral 1D domains rotated by 90° with respect to each other are found. The occurrence of only two or four building blocks and their 1D ("tire track" structure) or assemblies can be explained via 2D molecular self-organization symmetry operations [15], see also supporting information Chapter 4. The possible number of energetically favorable equivalent molecular orientations depends, on the one hand, on the molecular symmetry itself, on the other hand, on the symmetry of the substrate. Because the Cu(100) substrate has a fourfold symmetry and the DCzDCN molecule is not mirror symmetric in 2D, in total, eight energetically most favorable positions are possible. Therefore, for the DCzDCN molecule, no other than four- and two-fold assemblies can be expected.

Furthermore, despite the fact that the DCzDCN itself is chiral, it is remarkable that the 1D linear structures are always built from the same enantiomers. Mixed racemic domains, as were found for other chiral molecules on the Cu(100) surface at higher coverage [16], could not be identified for the 1D phases. This phase separation is clearly evident and can be seen, e.g., at the domain boundary between two homochiral phases A (M-(-)) and B (P-(+)), in Fig. 3d (yellow arrow). Just the domain structure of the "tire track" structure shows a significantly higher number of left-handed M-(-) chiral domains than P-(+) right-handed ones. Such behavior clearly indicates chiral resolution for the system DCzDCN/Cu(100). The two possible 90° orientations of the 1D structures are slightly different for the two enantiomers (rotated by 5°). From the STM image in Fig. 2c, we determine the two overall directions for the enantiomers as $[0\bar{1}6]$ and $[0\bar{6}\bar{1}]$ for the 1D M-(-) phase, and $[01\bar{4}]$, $[0\bar{4}\bar{1}]$ for the 1D P-(+) phase.

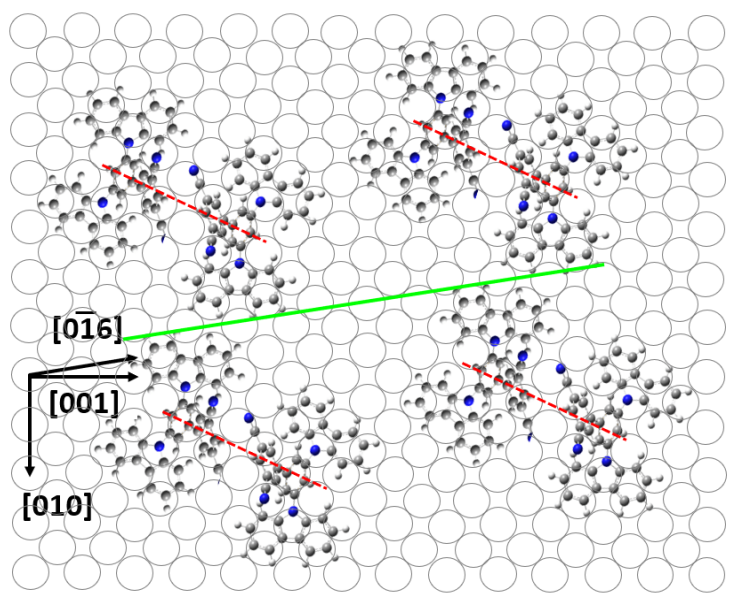

*Fig. 4: Tentative adsorption model showing the possible adsorption positions of the individual dimer pairs for the 1D homochiral M-(-) phase. The experimentally observed overall direction $[0\bar{1}6]$ of this phase is marked by a green line. For simplicity, the centers of the aromatic carbazole units are indicated by the red circular areas, and the phthalonitrile group is indicated by a blue circular area. The respective monomer orientation $[012]$ is indicated by red dashed lines.*

In a tentative model, the adsorption positions of the dimer building blocks within the 1D structure are constructed. We chose an approach that has been used for other aromatic systems on Cu(100) [16]. The starting point is the adsorption position of benzene on the hollow site of a Cu(100) surface, as indicated by DFT calculations [17]. The result is shown in Fig. 4. In the model, the observed [0$\bar{1}$6] direction of the 1D M-(-) phase can actually be reproduced. The individual dimer building blocks have a [012] orientation. The shown model perfectly fits with the observed 1D M-(-) phase in Fig. 6d. The two carbazole moieties interacting with the substrate are at an angle of 43° to each other. This results in an almost ideal arrangement with respect to the existing four-fold hollow adsorption sites of the Cu(100) substrate. Nevertheless, especially in view of the axial chirality of the DCzDCN molecule, the possibility of an STM-induced chirality change must also be considered. Such a tip-induced chirality change of individual molecules was recently demonstrated on a diphenyl carbene adsorbed on a copper surface. Similar to our system, the rotation about a C-C-C axis is partially fixed on the copper surface and causes a chirality change [18]. However, we have not observed any direct evidence for a tip-induced change in the chirality of individual DCzDCN molecules in our investigations.

**Photoemission spectroscopy**

The formation of molecule clusters on flat terraces in the sub-ML range indicates interaction between the DCzDCN molecules and the Cu(100) surface. To identify any kind of interaction between adsorbed molecules and metal substrates, the coverage dependence of the electronic structure upon adsorption is measured. Therefore, UPS and IPS spectra were subsequently acquired with increasing coverage to determine the molecular orbitals and possible hybrid interface states (HIS), as well as changes in the surface dipole ΔΦ. ΔΦ will change by any kind of charge transfer between substrate and molecule and/or charge redistribution at the organic/metal interface, and is, in addition, often a first indication of the orientation of polar molecules like DCzDCN on the surface.

**Ultraviolet photoemission spectroscopy/ Inverse photoemission spectroscopy**

The coverage dependence of UPS spectra is shown in Fig. 5. The spectrum for the clean copper surface is dominated by the intense Cu 3d peaks centered around -3 eV and the broad s-p plateau between -2 eV and the Fermi edge ($E_F$). The development of the work function Φ (for details, see supporting material, Fig. S4) and, therefore, ΔΦ, which we get from the changes of the secondary cut-off (Fig. 5 a), was evaluated with increasing evaporation time of DCzDCN up to 230 min, corresponding to a coverage >> 2ML and thus approximately bulk-like behavior.

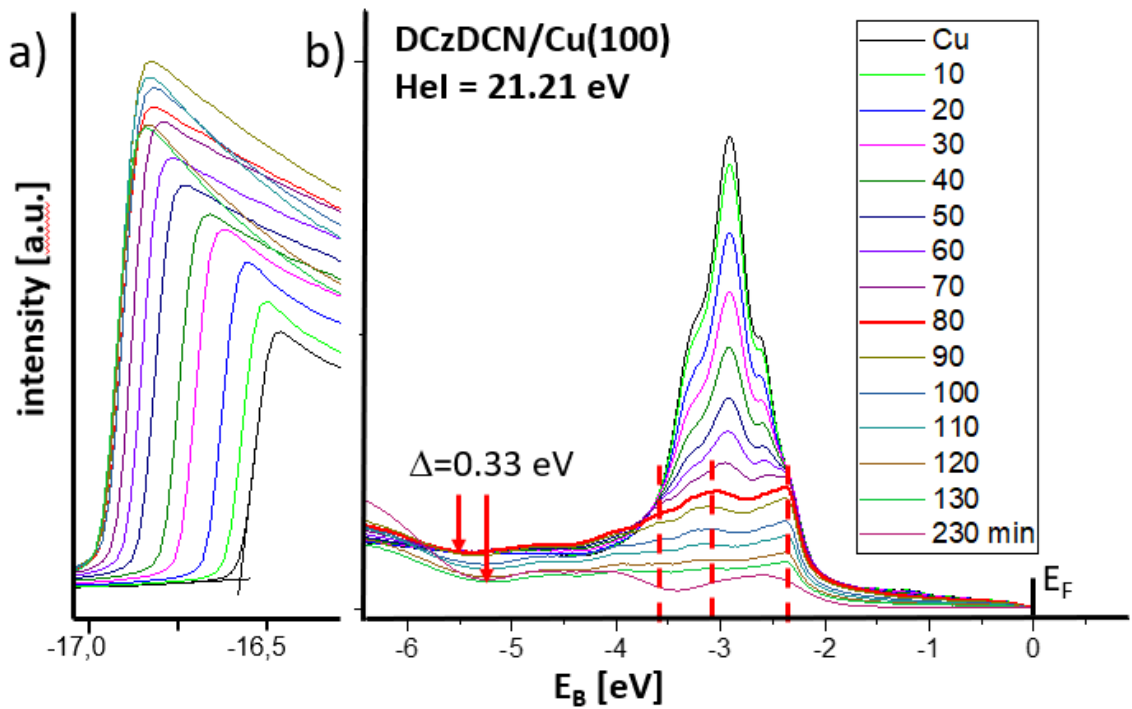

Fig. 5: HeI UPS spectra for increasing coverage (evaporation time) of the system DCzDCN/Cu(100): a) Secondary electron cut-off; b) valence region up to the Fermi level ($E_F$). Dashed lines mark states with irregular attenuation of intensity. The positions marked with arrows show the energetic shift of the spectra during the transition from monolayer to multilayer (for further details, see text).

The initial value for the work function of a clean Cu(100) surface is $\Phi = 4.65\pm0.05$ eV and corresponds very well with values found in the literature [19,20]. The evaluation of the work function (Fig. 5 a) and Fig. S4) with an increasing DCzDCN coverage shows a continuous decrease of $\Phi$ and, therefore, an increase of a positive surface dipole $\Delta\Phi$. Furthermore, the saturation behavior of $\Delta\Phi$ for an evaporation time of 80 min clearly indicates the formation of the first monolayer (ML) with a corresponding work function of $\Phi = 4.19$ eV. A typical value of 3.99 eV for higher coverage (230 min) in the multilayer region (>> 2ML) was further used as the reference value for the bulk work function $\Phi_{bulk}$ of DCzDCN in the energy level diagram shown later in Fig. 9.

There are three main causes for the change in the work function: The first effect is the general Pauli push-back effect, which is always present at a metallic surface in interaction with an organic molecule [21]. It will always decrease $\Phi$.

The second is the perpendicular part of the molecular dipole moment $\mu_0$ to the surface dipole (as is the case for DCzDCN, see Fig.1), which depends on the adsorption geometry of the molecules. A standing adsorption geometry leads either to a negative $\Delta\Phi$ in the case of adsorption via the carbazole units or to a positive $\Delta\Phi$ in the case of adsorption via the phthalonitrile moiety. From the STM results, we thus expect an increase in work function.

The third is the bond dipole (BD), which quantifies the charge rearrangements at the interface caused by any hybridization between an adsorbed molecule and the substrate. (Because of the metallic substrate, a purely integer charge transfer (CT) can be excluded because it is based on tunneling through insulating layers [22,23].) Therefore, the BD reflects the net charge transfer between the metal surface and the molecule, generated by the degree of hybridization. While the calculation of the real BD requires DFT calculations based on the knowledge of the hybridization of the molecular states with the mostly d-like states of the metal surface, the possibility of a charge rearrangement at such an organic/metallic interface can already be estimated via the chemical potentials of the substrate and the molecule, with the electronic charge transferred from the species with the higher chemical potential to the one with the lower chemical potential. This can be estimated in a simple approach by the so-called finite-difference approximation, which defines the chemical potential of the molecule as the negative of the average of the first ionization potential IP and the electron affinity EA, $\mu = -(IP+EA)/2$ [24]. The chemical potential of the Cu(100) substrate can thus be derived from $\Phi$, leading to $\mu(Cu) = -4.65$ eV. Measuring EA and IP of the >>2ML thick layer of DCzDCN by inverse photoelectron spectroscopy (IPES) and UPS, we found EA = 3.14 eV and IP = 5.59 eV. The chemical potential of pristine DCzDCN can then be estimated to $\mu(DCzDCN) = -4.39$ eV. Therefore, within the framework of this assessment, the direction of a CT will be from the DCzDCN to the Cu-substrate. Such a CT would decrease $\Delta\Phi$. Because we see an overall decrease of $\Delta\Phi$ (Fig. S4), Pauli repulsion and CT together more than counteract the effect of the orientation of the molecular dipole.

Looking at the development of the valence band structure in Fig. 5b, we can first see a shift of the parts of the UPS spectra by 0.33 eV, which correlates with the change from mono- to multilayer growth (red arrows in Fig. 5xa). In the main part of the UPS spectra in Fig. 5b, which is dominated by the Cu 3d peaks around -3 eV, we see unusual intensity damping behavior (marked by the red dashed lines) that is not consistent with a simple continuous attenuation of the Cu 3d states by scattering of the photoelectrons at the hybrid interface. Such an irregular damping behavior is just as typical for the appearance of purely molecular as for new hybrid interface states (HIS). To check the reason for this irregular damping, we analyzed the UPS and IPES spectra for a ML and a thicker layer >>2ML in detail (Fig. 6).

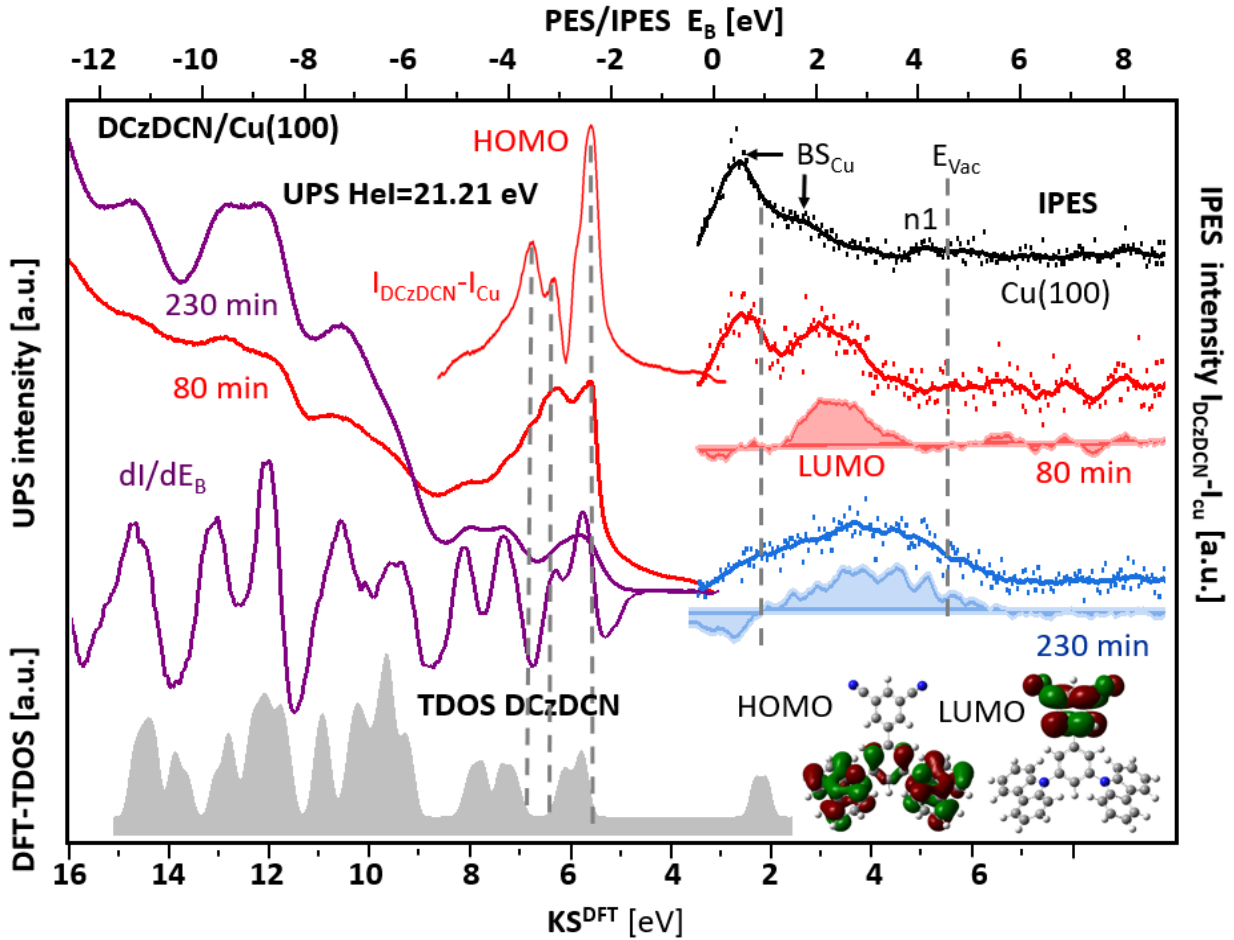

*Fig. 6: Detail spectra of the occupied (left part) and the unoccupied electronic structure of the DCzDCN/Cu(100) interface (80 min red) and a bulk-like layer of DCzDCN (230 min purple/blue). For more details, substrate-subtracted spectra were used in the case of the ML coverage (80 min) (inset) and the second derivative of the spectra for the bulk-like layer (230 min). For evaluation of the IPES data, an additional IPES spectrum of the pristine Cu(100) substrate is shown (black). The IPES spectra are also given as difference spectra. Additionally, a theoretical total DOS from DFT calculations for the single molecule DCzDCN up to the LUMO-2 state (grey part) and the resulting molecular orbitals for the HOMO and the LUMO are also shown (for further details, see text).*

For this purpose, we chose the method of difference spectra due to the interfering 3d intensities of the substrate in the spectrum for a ML. Thus, by subtracting a normalized substrate background, the pure molecular or hybrid states can be highlighted. Spectra of thicker layers, moreover, are often washed out with respect to the finer structures. Therefore, the second derivative of the spectra was used to make these details visible again. In order to obtain a complete picture of the electronic structure of the system DCzDCN/Cu(100 and bulk-like DCzDCN, the unoccupied states were determined by using IPES (Inverse Photoemission Spectroscopy) on these two layers. Also for the IPES spectra, we subtracted the Cu substrate background. The resulting data sets are shown in Fig. 6. By comparing the total density of states (TDOS Fig. 6 shaded gray), obtained by DFT calculations (see also Fig. S1 supporting information) on the single molecule, with the UPS spectrum of the bulk-like layer (Fig. 6 purple), a very good agreement and thus assignment to single molecular states could be achieved. For this layer, the signature of the HOMO can be identified at -2.3 eV and the HOMO onset at 2.0 eV. Comparing the spectra for the thicker layer with the ML spectra, it can be seen that the peak associated with the HOMO is slightly shifted towards $E_F$. Also, the peak structure between -4 and -3 eV changes especially through the appearance of an additional peak structure at -3.5 eV, which cannot be identified in the bulk-like structure. Also noticeable is an apparent slight increase in the density of states for the ML in the region between 2.0 eV and the Fermi edge. This region is determined only by the substrate's flat s-p band, so a subtraction of the substrate background can be omitted here. Fig. 7 clearly shows that a simple intensity decrease occurs only directly at the edge of the d-states of the substrate. In the entire other region, a significant increase in the density of states is found up to 0.3 eV below $E_F$. Despite its quite weak intensity, such an increase of a DOS signature in the band gap region of the undisturbed molecule is a clear indication of the formation of a HIS between the DCzDCN and the Cu surface.

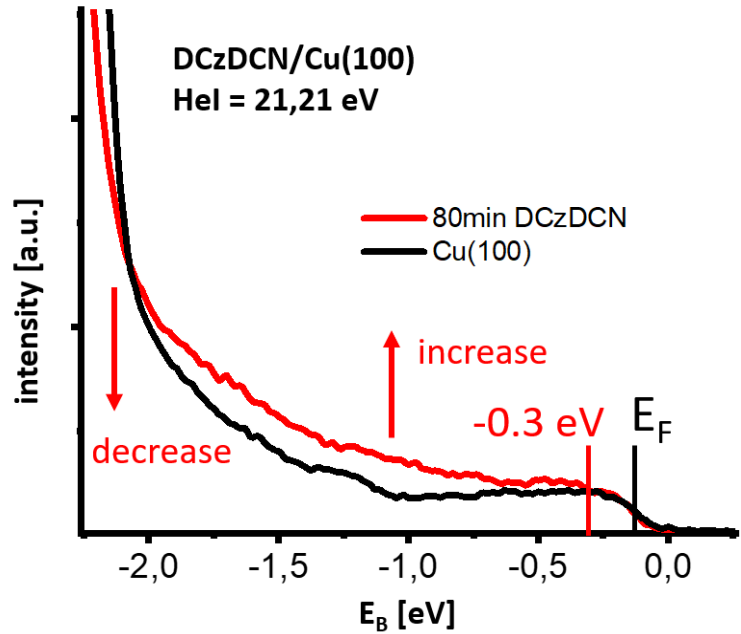

*Fig. 7: Detail spectra of the valence band region around the Fermi level for 1 ML. A clearly increasing intensity after adsorption of the DCzDCN indicates the formation of HIS in this region.*

For the unoccupied states, unfortunately, the IPES spectra are also dominated by the substrate-related states, as one can see for the pristine Cu(100) IPES spectra in Fig. 6 (black). The main peak structure at 0.4 eV is related to a p bulk-band transition ($BS_{Cu}$) [25]. The second structure, S1, is attributed to the surface resonance of crystal-induced states [26]. Because the S1 feature is susceptible to adsorbates, it is strongly quenched. The step-like feature n1 is attributed to the image potential state pinned to the vacuum level. For the ML, the feature corresponding to the LUMO at 2 eV can be seen, while the Cu bulk band transition (B) and the Cu surface resonance (S1) are clearly quenched. From the difference spectra (red dotted line with the red filled area under the curve), the onset of the LUMO orbitals at 1.1 eV can be determined. For the bulk-like layer (230 min deposition, blue dotted line with the blue filled area under the curve), the onset of the LUMO orbitals shift toward the Fermi level down to 0.8 eV, and the whole peak structure is significantly extended. Thus, the data from the UPS and IPES analysis clearly show that common hybrid interface states between the DCzDCN and the Cu surface build up in the first ML. Because of the existence of a HIS, the appearance of a BD is the consequence.

**Scanning Tunneling Spectroscopy**

Since the determination of electronic densities of states by UPS always represents only an integral snapshot over a larger surface area, the region up to 3 eV below the Fermi level was investigated by means of tunneling spectroscopy (STS). As a single molecule spectroscopy, STS allows direct access to electronic states with submolecular resolution. Thus, the role of individual molecular regions in the framework of the total DOS can be better identified.

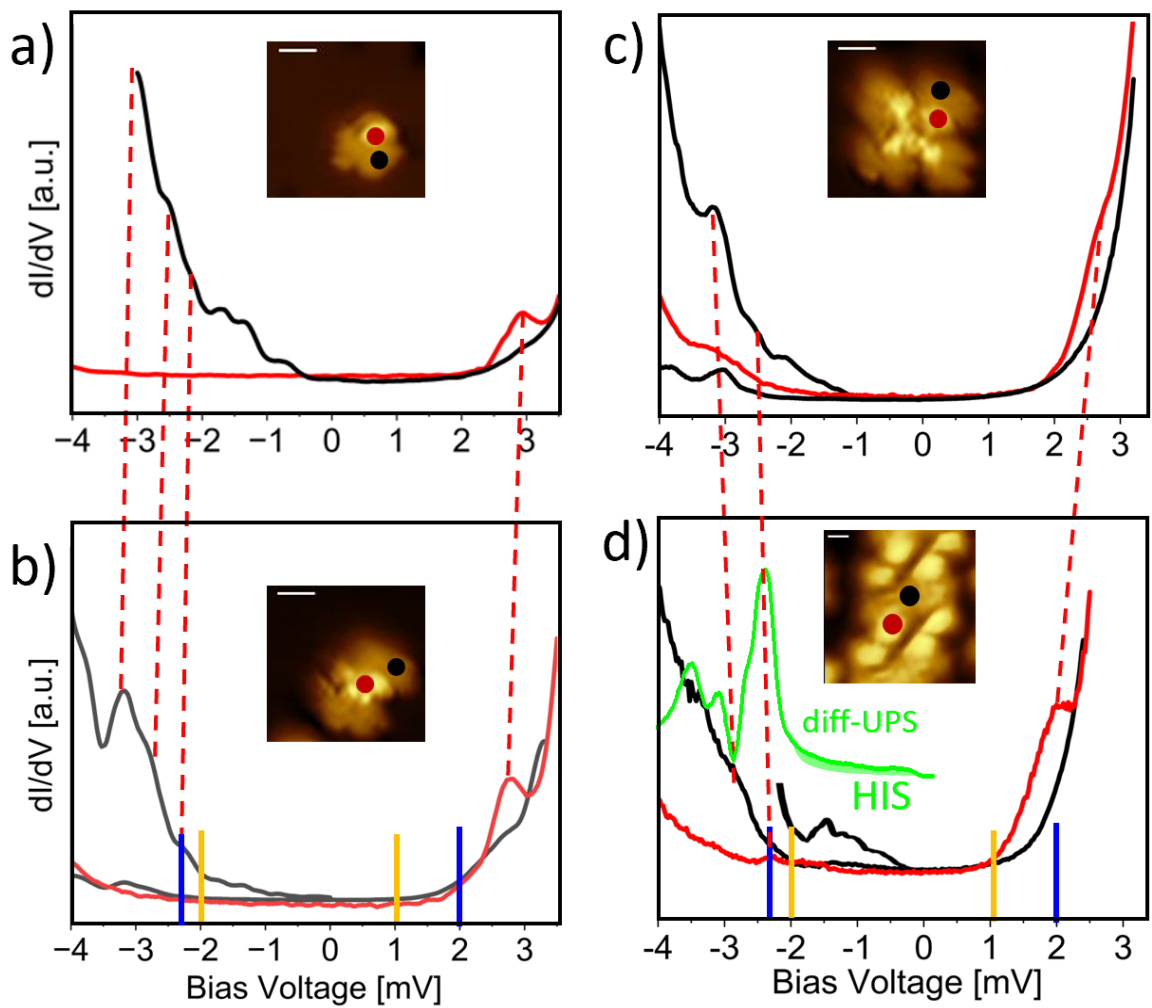

*Fig. 8: dI/dV point spectroscopy curves (STS spectra). The point spectroscopy curves and their related locations (shown as spots in the corresponding inset on high-resolution STM topography images) are marked by black and red curves. Color code: red = phthalonitrile acceptor moiety, black = carbazole donor moiety. a) for a monomer, b) a dimer, c) a tetramer, and d) the 1D "tire track" structure. Inset: UPS spectra (green) showing HIS structure (green area); similar structures are connected by red dashed lines. For comparison, the peaks (blue line) and the onset (orange line) of the HOMO and LUMO from UPS/IPES measurements are shown in b) and d). Stabilization parameters: 50 pA and 3.5 V for (a,b), 50 pA and 3 V for (c,d).*

In Fig. 8, dI/dV point spectroscopy curves (STS spectra) for single molecules up to the 1D "tire track" structure are shown. Since we know the orientation of DCzDCN molecules from the STM topography images, we extended our point measurements to two distinct positions on the molecule. The first position is located on the carbazole moiety in direct contact with the Cu substrate. The second position is on the phthalonitrile/biphenyl moiety, which protrudes out of the surface. As we know from the DFT calculations, the LUMO region is mostly related to the phthalonitrile/biphenyl moiety, the HOMO region is mostly at the carbazole moiety (Fig. S1 supporting information). This is reflected by the dI/dV curves in Fig. 8.

We are aware that the validity of dI/dV measurements is critical at higher bias voltages. The validity of the commonly used Tersoff-Hamann model [27] may already be exceeded at the voltages used. However, the PES/IPES measurements provide independent information about the surface/molecular states. In this combination, assignments of states clearly beyond 1 eV above or below $E_F$ are possible [28]. Thus, we take our PES data as a hint to assign the STS peaks to the respective molecular orbitals. For this purpose, the PES peak position (blue lines) and the onset (orange line) of the HOMO and LUMO are marked. From this, we assign the structure at 3 V in the region of the unoccupied states to the LUMO on the acceptor part (phthalonitrile/biphenyl moiety) of the molecule. In the region of occupied states, the HOMO at -2.3 V is located at the donor part (carbazole moiety) of the molecule. These structures can be found in all stages, from the monomer to the 1D structure. Going from the monomer to the dimer, all the structures in the spectra for the occupied as well as the unoccupied states are only slightly shifted to higher binding energy. In the transition to the tetramer and the 1D "tire track" structure, the states of the occupied and unoccupied regions each move toward the Fermi level. This effect can be attributed to a polarization effect by the surrounding molecules, i.e., the influence of the transient local electrostatic environment (screening) induced as a response to instantaneous localized charge added or removed during tunneling spectroscopy. This effect is well known in PES experiments and has recently also been observed by STS performed on molecular adsorbates [29,30]. The stabilization reduces the measured ionization potential (observed as an upshift of HOMO energy) and increases the electron affinity (observed as a downshift of LUMO energy [31]. Because of this effect and the fact that the spectra from the 1D structure in Fig. 8d were made on a comparable surface coverage as the PES shown in the inset and the IPES in Fig. 6, the comparison of the peaks and the onset energies for the LUMO and the HOMO between the STS spectra and the values from the PES/are allowed. Fig. 8d also shows that in the region between the HOMO onset at -2 V and the Fermi level, densities of states are also found as in the PES spectrum. Since these densities of states are relatively small, the question arises whether these stem purely from tunneling from the substrate into the gap of the organic layer or are genuine new HIS structures. If we consider the energetic position of the first occupied molecular orbitals, they are in the direct range of the Cu d states. This is a typical situation, which is predestined for a resonant interaction of organic MOs in the framework of an adsorption model, according to Newns-Anderson-Grimley [32,33]. Here, the MOs of the organic adsorbate coupled with the Cu d states would lead to exactly such a new density of states, which is strongly broadened in the direction of $E_F$, i.e., a HIS (see supporting material Fig. S5). Thus, from the comparison between the STS and PES/IPES data, the energy level diagram in Fig. 9 can be constructed.

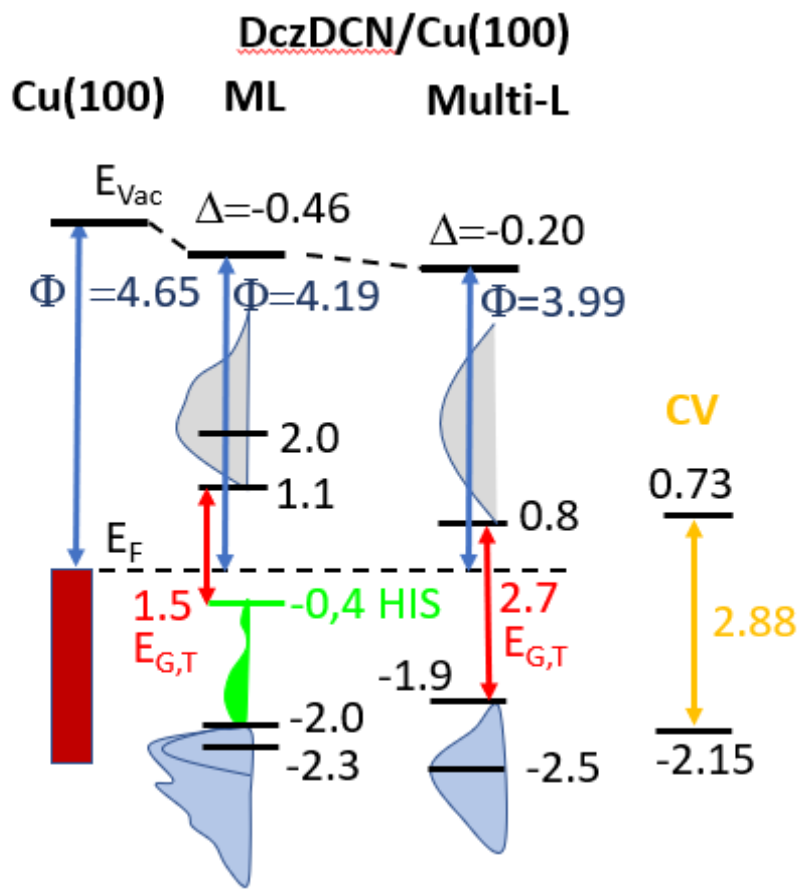

*Fig. 9: Energy level diagram and transport gap $E_{G,T}$, as it results from the analysis of the UPS/IPES and STS data for a ML and UPS/IPES for a multilayer. The onset for the HIS region (green area) starts at -0.4 eV. Also shown are the HOMO and LUMO levels from the literature [y,] calculated from cyclic voltammetric measurements (CV, orange).*

The HOMO and LUMO levels, calculated from the reduction and oxidation potentials measured by cyclic voltammetry (CV), are already described in the literature [11] for the pure DCzDCN. This can be compared with the energy levels found for the multilayer system (Multi-L) (see Fig. 9). A transport gap of $E_{TG}$ = 2.7 eV, determined by the onsets of the occupied and unoccupied states for the respective systems, is found for the condensed DCzDCN. The slight deviations of the values for the multilayer compared to the values from the CV measurements come, first of all, from a different polarization behavior of the individual molecules in solution compared to the condensed phase.

For the DCzDCN/Cu(100) interface (ML spectra), the values change. The formation of a HIS by hybridization of the carbazole units of the DCzDCN with the d-states of the Cu substrate leads to a significant change of -0.9 eV in the $E_{TG}$ and furthermore to a modification of the donor properties of the carbazolic part of the molecule itself. All this will directly affect the donor-acceptor interaction and, therefore, the exciting states, which determine the optical behavior of the adsorbed molecule compared to the single one.

**Summary**

In this study, we investigated the interaction of the TADF active donor-acceptor molecule DCzDCN with a Cu(100) surface. By analyzing STM images with sub-molecular resolution, an exclusively standing orientation of the molecules with their phthalonitrile acceptor moiety protruding from the surface plane is found. This adsorption geometry enables the carbazole moiety, the donor part of the molecule, to interact with the Cu substrate. Because of an axial chirality of the DCzDCN molecules, two enantiomers exist (*M*-(-), *P*-(+)). Adsorption at lower coverage on the Cu(100) surface leads to the formation of homochiral building blocks of di- and above all tetramers. At higher coverage, instead of the previously dominant tetramers, individual dimer building blocks are found that assemble into homochiral "tire track"-like 1D structures. For each of the two groups, we found two mutually perpendicular preferred directions, with slightly different orientation and slight preference for the M-(-) phase. The driving force behind all of these self-ordering processes is, on the one hand, an interplay between the fourfold symmetry of the Cu(100) surface and the arrangement of the two carbazole units relative to each other, which is favorable, especially for this substrate. On the other hand, a strong coulomb force triggered interaction between the acceptor moiety of neighboring molecules enables the actual formation of di- and tetramers. The observation of only homochiral 1D structures in the ML range with a slight excess of M-(-) dimer-determined phases suggests that chiral resolution occurs at the DCzDCN / Cu(100) interface. Through a combined in situ analysis of the electronic structure of the DCzDCN/Cu(100) interface by both the integral measurement methods UPS/IPES and the single-molecule spectroscopy by STS, the formation of a hybrid interface state at this interface could be identified. The hybridization can be explained in the framework of a Newns-Anderson-Grimley adsorption model via the interaction of the molecular states in the HOMO region of the DCzDCN with the d-states of the Cu substrate. The HIS is mainly localized at the donor part of the molecule. The new density of states reduces the transport gap of the adsorbed molecule by 0.9 eV compared to the value for the free molecule. It turns out that in a ML of DCzDCN molecules on a Cu(100) surface, new bulk-unlike structures with modified electronic properties were formed because of chiral resolution at the interface. The question of effects on the excited molecular states as well as the directly related optical properties of the system DCzDCN/Cu(100) are therefore exciting questions to be answered in case of using such a system for further, maybe new (opto-) electronics applications. In further studies, this system will, in addition, be studied in the context of a possible chemically induced spin-polarization.

**Methods**

Experiments were performed by using an ultrahigh vacuum multi-chamber system (PREVAC, Rogów, Poland), allowing sample transport among preparation and analysis chambers under a base pressure below $2 \cdot 10^{-10}$ mbar.

**Sample preparation**

Single crystal Cu(100) (MaTecK, Jülich, 99.999% purity, <0.1° orientation accuracy, <0.03 μm surface roughness) was cleaned by repeated cycles of Ar+ ion bombardment at 1 kV and subsequent annealing to 500°C. The quality of the crystal surface was examined by using photoelectron spectroscopy and scanning tunneling microscopy. DCzDCN molecules (Luminescence Technology Corp. New Taipei City, Taiwan, sublimated 99%) were thermally evaporated at 240°C and deposited under UHV conditions onto the clean Cu(100) single crystal surface kept at room temperature. The deposition rate was controlled by a using quartz microbalance and by a comparison of STM/UPS measurements.

**Photoemission measurements**

Photoemission studies were performed with a hemispherical energy analyzer PHOIBOS 150. (SPECS Surface Nano Analysis GmbH, Berlin) For ultraviolet photoelectron spectroscopy, a conventional gas discharge lamp was used with an excitation energy of 21.22 eV for the He I line. During photoelectron collection, the chamber pressure was kept at $10^{-10}$ mbar, and on samples -6 V bias was applied to ensure accurate determination of the low kinetic energy cut-off. Work function has been determined from secondary electron cut-off in normal emission. Inverse photoelectron spectroscopy measurements were performed using a homemade setup equipped with an EGPS-2B electron source (Kimball Physics, Wilton, USA) with a barium oxide cathode. The sample was moved after each scan to avoid film damage.

**STM/STS measurements**

Scanning tunneling microscopy and spectroscopy in situ measurements were performed by using a Pan-Scan microscope (RHK Technology, Troy, USA) operating with a closed cycle cryostat, enabling measurements at 9.5K (on a sample). Tungsten tips were produced by electrochemical etching, while further in situ annealing and sputtering cycles allowed oxide film removal. The STM images were acquired in constant current mode with a relatively high bias voltage (up to 3V) and low tunneling current (50 pA) and then processed by Gwyddion open-source SPM analysis software. STS curves were acquired by using the internal RHK lock-in amplifier at 4 kHz modulation frequency and 50 meV modulation amplitude. The relatively high modulation amplitude for a wide range (-4V;3.5V) and relatively strong shifts provide sufficient resolution while ensuring a high signal for spectra collection. Curves were discarded if the current jump indicated alteration or reorientation of a molecule.

**TDOS and single molecular geometry calculations**

The optimized single molecular geometry calculations for the DCzDCN were carried out using DFT with a Perdew-Burke-Ernzerhof (PBE) functional and the def2-TZVPP basis set, as implemented under GAUSSIAN03 [34]. For the electrostatic potential map ESP, an isoval of 0.0004 was used. For the TDOS spectra, each Kohn Sham eigenstate from the single molecular ground state calculations was broadened by a Gaussian function of 0.2 eV FWHM.

AUTHOR INFORMATION

**Corresponding Authors**

S. Lach

E-mail: lach@physik.uni-kl.de

RPTU Kaiserslautern-Landau

Erwin Schrödinger Str. 56

D-67663 Kaiserslautern

R. Ranecki

E-mail: ranecki@physik.uni-kl.de

0631 205 4600

RPTU Kaiserslautern-Landau

Erwin Schrödinger Str. 56

D-67663 Kaiserslautern

**Funding Sources**

This work was funded by the Deutsche Forschungsgemeinschaft (DFG, German Research Foundation) - TRR 173-268565370 Spin + X (projects No. B05).

**Supporting information includes:**

TDOS and MO-symmetry by single molecule DFT calculations

Electrostatic potential map ESP

Structural assignment of individual assemblies for sub-monolayer coverage

Symmetry considerations for the system DCzDCN-Cu(100)

Coverage dependence of the work function

HIS formation following the Newns-Anderson-Grimley adsorption model

ACKNOWLEDGMENT

This funding by the Deutsche Forschungsgemeinschaft (DFG, German Research Foundation) - TRR 173-268565370 Spin + X (project No. B05) is greatly appreciated.

**Supporting Information:**

# Self-assembly of the chiral donor-acceptor molecule DCzDCN on Cu(100)

*Robert Ranecki, Benedikt Baumann, Stefan Lach, Christiane Ziegler*

Department of Physics and Research Center OPTIMAS, Erwin-Schroedinger-Str. 56,

67663 Kaiserslautern (Germany)

1. ***TDOS and MO-symmetry by single molecule DFT calculations***

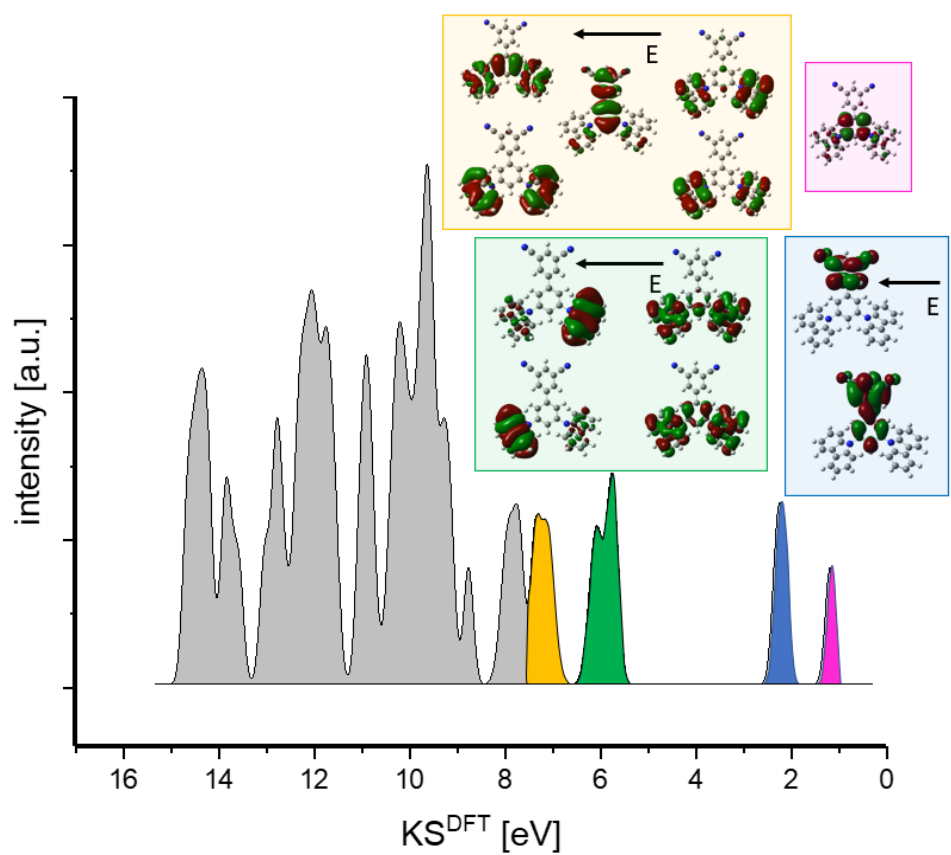

*Fig. S1: Calculated TDOS for a single DCzDCN molecule using PBE and the def2-TZVPP basis set. Also shown are the most important molecular boundary orbitals and their assignment to the TDOS.*

## 2. Electrostatic potential map ESP

The image in Fig. S2 shows the electrostatic potential map ESP with isosurface (isoval=0.0004) calculated by DFT using PBE and the def2-TZVPP basis set, showing the charge density distributions for DCzDCN (ESP contours are color-coded from blue (positive) to red (negative) potentials.

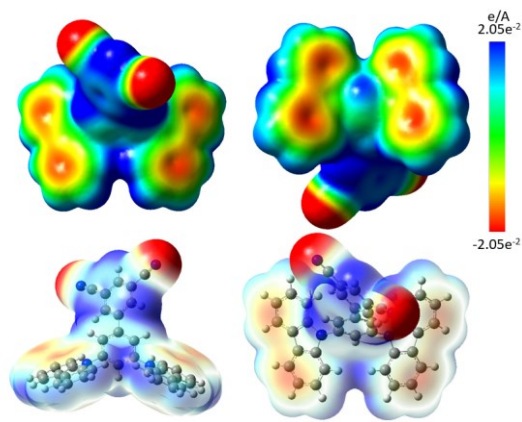

Fig. S2: Electrostatic potential map ESP with isosurface (isoval=0.0004) for a single DCzDCN molecule (color-code blue = positive, red =negative)

## 3. Structural assignment of individual assemblies for sub-monolayer coverage

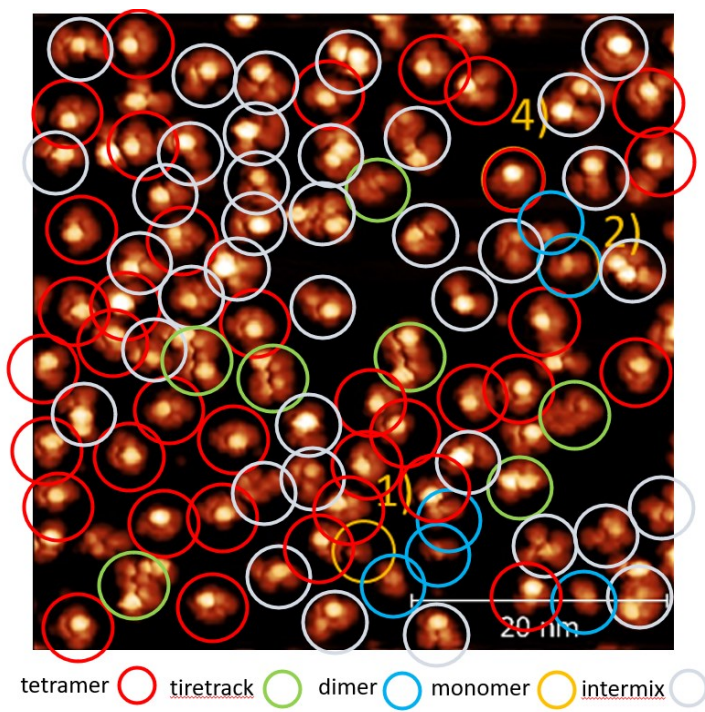

Fig. S3: Structural assignment of individual assemblies with submonolayer coverage corresponding to Fig. 2b in the main paper.

We have investigated the different assemblies of DCzDCN in the sub-monolayer region. This showed that 72% of all unambiguously assignable structures are tetramers and a roughly equal distribution of 14% dimeric and 12% building "tire track" structures.

4. *Symmetry considerations for the system DCzDCN-Cu(100)*

The found di-/tetramers, as well as the 1D "tire track" structure of dimer pairs, can be explained via 2D molecular self-organization symmetry operations [1]. The possible number of energetically favorable equivalent molecular orientations depends, on the one hand, on the molecular symmetry but also on the symmetry of the substrate. The number of observable molecular adsorbate orientations $N_{orient}$ is related to the number of elements of the substrate rotational symmetry group $C_{i-surf}$ and the number of elements of the largest subgroup $C'_{k-Mol}$ of the molecular rotational symmetry group $C_{j-mol}$ being also a subgroup of $C_{i-Surf}$. Considering the set of rotations on the four-fold symmetric Cu(100) substrate, the set will be:

S = {rotate 0°, rotate 90°, rotate 180°, rotate 270°} indicating the geometric operation (in our case, rotation by 90°)

$C_{i-Surf} = C_{4-Surf}$ = {rotate 0° = 360°, rotate 90°, rotate 180°, rotate 270°} = 4

Because of the chiral character of the DCzDCN molecule, there is only one possible geometric operation S={rotate 360°}; therefore, we get $C'_{k-Mol} = 1$.

Then, for the system DCzDCN/Cu(100), we find

$$N_{Orient} = \frac{C_{4-surf}}{C'_{1-Mol}} = 4$$

In case the molecule is not mirror-symmetric in 2D, which is the case for the DCzDCN, the number of energetically most favorable positions doubles. Therefore, for the DCzDCN molecule, no other than four- and twofold assemblies can be expected [1].

## 5. *Coverage dependence of the work function*

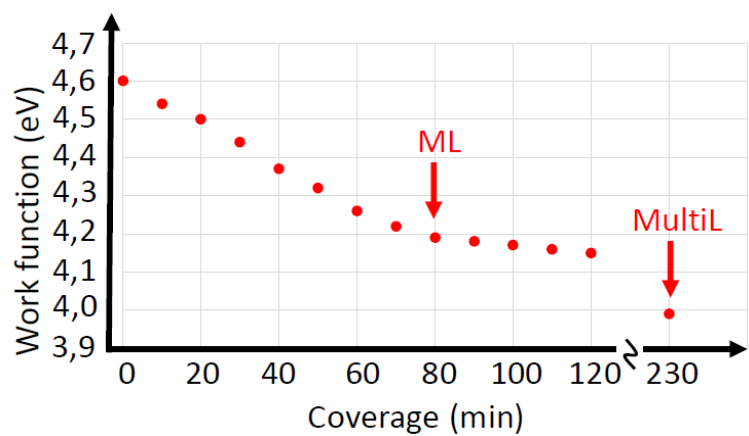

Fig. S4: *Coverage dependence of the work function Φ for the system DCzDCN/Cu(100)*

The work function Φ shows a uniformly decreasing value until saturation at 4.19 eV, indicating the formation of the first ML. A typical value of 3.99 eV for higher coverage in the multilayer region (>> 2ML) was found and further used as the reference value for the bulk work function of DCzDCN.

## 6. HIS formation following the Newns-Anderson-Grimley adsorption model

A schematic comparison of the effects on the electronic structure of a chemisorbed molecule concerning the interaction with sp- and/or d-bands of a metallic substrate, following the Newns-Anderson-Grimley adsorption model is shown in Fig. S5 [2,3].

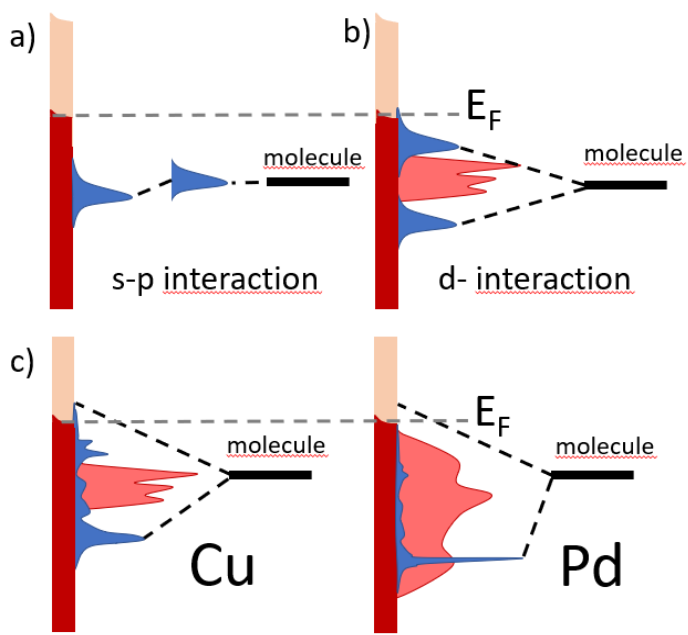

*Fig. S5: Coupling of molecular states to the d-band structure of a metal substrate following the Newns-Anderson-Grimley adsorption model (partly adapted from [4]). a) Coupling of a discrete organic state to the sp-band of a metal substrate, b) Coupling to the d-band structure of a metal substrate, c) Resulting adsorbate DOS depending on the width and energetic position of the d-DOS on the d-band structure of the metallic substrate.*